# On a Hybrid Preamble/Soft-Output Demapper Approach for Time Synchronization for IEEE 802.15.6 Narrowband WBAN

**Imen Nasr**[1,2], **Leila Najjar Atallah**[2], **Sofiane Cherif**[2], **Jianxio Yang**[1], **Kunlun Wang**[3]
[1]Unité d'Informatique et d'Ingénierie des Systèmes (U2IS), Ecole Nationale Supérieure de Techniques Avancées (ENSTA ParisTech), Palaiseau, France.
[2]Communication, Signal and Image (COSIM) Laboratory, Higher School of Communications of Tunis (Sup'Com), Ariana, Tunisia
[3]Network Coding and Transmission Laboratory, Shanghai Jiao Tong University, Shanghai, China

**Abstract**: In this paper, we present a maximum likelihood (ML) based time synchronization algorithm for Wireless Body Area Networks (WBAN). The proposed technique takes advantage of soft information retrieved from the soft demapper for the time delay estimation. This algorithm has a low complexity and is adapted to the frame structure specified by the IEEE 802.15.6 standard [1] for the narrowband systems. Simulation results have shown good performance which approach the theoretical mean square error limit bound represented by the Cramer Rao Bound (CRB).

## I. INTRODUCTION

WBAN are currently trying to meet new markets demand in diverse areas such as security, health care, gaming, smart clothes... These WBANs constitute a central component of the Internet of Things (IoT). Individuals, with sensors and actuators, can indeed interact more effectively with their environment, while ensuring global connectivity via the mobile internet. These wearable networks are based in part on emerging wireless technologies with very low power consumption such as narrowband systems which were proposed as a part of the IEEE 802.15.6 WBAN standard [1].

A WBAN is operating in a heterogeneous environment where different wireless technologies coexist in the same frequency band. Thus, in order to insure reliable data transmission, time synchronization is a crucial task in a WBAN receiver providing robustness to noise, interference and contention. As time synchronization is processed at front-end, a bad synchronization degrades the whole receiver.

Both Data Aided (DA) and Non Data Aided (NDA) time delay estimation techniques have been employed in real systems such as in [2] and [3]. Even if DA techniques lead to the best achievable performance, one needs to look for other solutions not requiring the transmission of a pilot signal in order to enhance the spectral efficiency and to save power. To deal with this problem, NDA time recovery techniques are implemented using only the received signal. Unfortunately, compared to DA techniques, the system performance is degraded especially at low Signal to Noise Ratio (SNR) values. To overcome such disadvantages, with the development of channel coding techniques and iterative receivers [4-8], Code Aided (CA) synchronization algorithms [9] have been developed to enhance time delay estimation by re-injecting the soft output of the decoded signal to the synchronizer. In this case, no pilot signal is needed. Also, the timing synchronizer and the channel decoder can improve each other progressively by exchanging information [10], [11]. However, this technique cannot be held in a WBAN context due to its implementation complexity. Therefore, we propose in this paper to introduce the soft output of the demapper in the synchronization process to ameliorate the performance of the time delay estimator in the NDA mode. Such approach allows processing an accurate synchronization in the absence of pilot signals. It also allows avoiding the transmission of such pilot signals, representing an expensive overhear in terms of power consumption and spectrum overhead.

This paper is organized as follows. In section II, the IEEE 802.15.6 standard specifications for the narrowband WBAN operating in 2.4GHz frequency band are presented. In section III, the soft time synchronization algorithm is proposed. Simulation results are provided in section IV and validate our analysis. The last section concludes our work.

## II. WBAN STANDARD SPECIFICATIONS

A WBAN in general and in the case of medical applications in particular, is collecting critical information from different body parts. Therefore, reliability and latency of data transmission are two extremely important parameters to be taken into account in the design of the physical (PHY) and medium access control (MAC) layers.

On the other hand, WBAN are wearable sensors, which are implanted in the body or fixed on the body surface. This is why, power usage is a constraint of utmost importance for the system design which can be optimized at the PHY and MAC layers. The MAC layer can insure efficient power management by a good choice of packet scheduling, channel access and signalling techniques. The PHY layer enhances power saving by increasing the successful packet transmission probability or in other terms decreasing the bit error rate (BER). This can be achieved by adopting adaptive coding and modulation techniques to deal with the transmission channel problems.

Despite the critical needs to be power efficient and reliable, the complexity of both the PHY and MAC layers must be as low as possible, since sensors have limited calculus capability. The narrowband (NB) PHY layer specifications of the IEEE 802.15.6 standard were specifically designed to meet this low complexity requirement and we now describe the standard. TABLE I summarizes the coding and modulation techniques specified by the standard for NB systems in the frequency band from 2400 MHz to 2483.5 MHz.

**Table I. Modulation and Coding Parameters**

| Packet Component | Modulation | Symbol Rate(kbps) | Code Rate (k/n) | Spreading Factor | Data Rate(kbps) |
|---|---|---|---|---|---|
| PLCP Header | $\pi/2$-DBPSK | 600 | 19/31 | 4 | 91.9 |
| PSDU | $\pi/2$-DBPSK | 600 | 51/63 | 4 | 121.4 |
| PSDU | $\pi/2$-DBPSK | 600 | 51/63 | 2 | 242.9 |
| PSDU | $\pi/2$-DBPSK | 600 | 51/63 | 1 | 485.7 |
| PSDU | $\pi/4$-DQPSK | 600 | 51/63 | 1 | 971.4 |

This system uses systematic BCH(63, 51, t = 2) encoder or its shortened version BCH(31, 19, t = 2) where t = 2 refers to the number of errors that can be corrected after decoding. A BCH code is chosen for NB WBAN for its easy-low-complexity decoding. Redundant bits are added to the transmitted packet for a possible error correction at the receiver. However, the limited number of additional bits does not lead to a severe increase of the power budget. We also note that a spreading factor depending on the information data rate, the modulation and the coding parameters, is introduced in order to ensure a constant symbol rate. The spreading factor is the number of times that an encoded bit is repeated. In bad channel conditions, it is preferable to use a low order modulation.

The main issue of the logical layer is to provide a method for transforming a physical-layer service data unit (PSDU) into a physical-layer protocol data unit (PPDU). As shown in Fig. 1, in order to create the PPDU, the PSDU is preceded during the transmission by a physical-layer preamble and a physical layer header. The physical-layer preamble and header are used in the demodulation, decoding and delivery of the PSDU at the receiver.

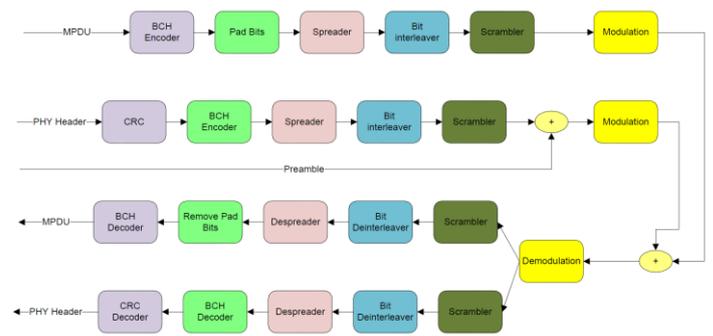

**Fig.1 Data Flow Structure**

Fig. 2 shows the format of the PPDU. It is composed of the physical layer convergence protocol (PLCP) preamble, the PLCP header and the PSDU, listed in the order of transmission. The PLCP preamble is the first component of the PPDU. Two unique preambles are defined in order to reduce false alarms coming from the networks operating on adjacent channels. Each preamble is constructed by concatenating a 63-length m-sequence with the following 27 bits extension sequence 010101010101101101101101101. Therefore, the preamble length is equal to 90 bits. The 63-length sequence is used for packet detection, coarse-timing synchronization, and carrier-offset recovery, while the latter sequence is used in fine timing synchronization.

The PLCP header is the second main component of the PPDU. This component conveys the necessary information about the PHY parameters used in the decoding of the PSDU. The 31-bit-PLCP header can be further decomposed into a PHY header, a header check sequence (HCS), and shortened BCH parity bits. Some information about the bit rate, the PSDU length, the burst mode and the scrambler can be retrieved from the PHY header. The HCS and BCH parity bits are added in order to improve the robustness of the PLCP header.

The PSDU is the last component of the PPDU. It is composed of the MAC header, the MAC frame body and the frame check sequence (FCS). The MAC header is composed of information proper to the MAC layer, such as, the transmitter and receiver ID, the WBAN ID and the frame type. The PSDU can then be encoded, spread and interleaved before being scrambled. The interleaver decreases the correlation between the transmitted symbols and the scrambler aims at having the expected power spectral density. When transmitting the packet, the PLCP preamble is sent first, followed by the PLCP header and finally the PSDU.

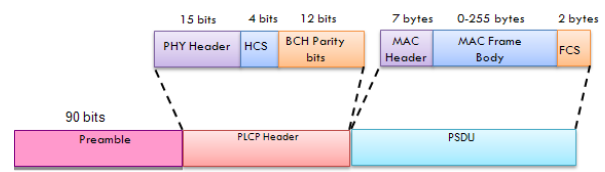

**Fig.2 Standard PPDU Structure**

The constellation mapper operates on the binary bit stream $b(n)$, which is the concatenation of the PLCP preamble, the PLCP header, and the PSDU. $b(n)$ is mapped onto one of three rotated and differentially encoded constellations: $\pi/2$-DBPSK, $\pi/4$-DQPSK and $\pi/8$-D8PSK. The mapping of the bit stream $b(n)$, $n = 1, ..,N$, onto a corresponding complex-values sequence $a(k)$, $k = 1, ..,N/log_2(M)$ is performed according to the following equation

$$a(k) = a(k-1)\exp(j\varphi_k), \quad (1)$$

where $a(0) = \exp(j\pi/2)$, $N$ is the total number of transmitted bits, $M$ is the constellation size and $\varphi_k$ is the phase change. Differential mapping is used to avoid the phase recovery and lower down the receiver complexity.

The symbol mapping depends on the chosen frequency bands of operation and the data rate (refer to Table I). In our subsequent simulations, we will consider a WBAN system operating in the band from 2400 MHz to 2483.5 MHz which, as in the standard, uses pi/2 -DBPSK constellation mapping for the preamble and the PLCP header and either 2 -DBPSK or 4 -DQPSK for the PSDU.

Depending on the selected mapping technique, $\varphi_k$ takes values according to TABLE II and III.

**Table II. $\pi/2$-DBPSK**

| $b(k)$ | $\varphi_k$ | $a_k a_{k-1}^*$ |
|---|---|---|
| 0 | $\pi/2$ | $\exp(j\ \pi/2)$ |
| 1 | $-\pi/2$ | $\exp(-j\ \pi/2)$ |

**Table III. $\pi/4$-DQPSK Mapping**

| $b(2k)$ | $b(2k+1)$ | $\varphi_k$ | $a_k a_{k-1}^*$ |
|---|---|---|---|
| 0 | 0 | $\pi/4$ | $\exp(j\ \pi/4)$ |
| 0 | 1 | $3\pi/4$ | $\exp(j\ 3\pi/4)$ |
| 1 | 0 | $7\pi/4$ | $\exp(j\ 7\pi/4)$ |
| 1 | 1 | $5\pi/4$ | $\exp(j\ 5\pi/4)$ |

We note that $a_k a_{k-1}^*$ can be seen as an imaginary binary modulated signal in the case of a $\pi/2$-DBPSK mapping and a quaternary modulated signal in the case of a $\pi/4$-DQPSK mapping.

Based on the above low-complexity PHY Layer specifications, we are going to develop in the next paragraph our adaptive timing recovery algorithm. It combines a DA estimation technique relying on the preamble known sequence and a soft technique exploiting the soft information coming from the demapper when the received sequence belongs to the PLCP header or the PSDU. To the best of our knowledge, no timing synchronization algorithms have been developed yet for NB WBAN.

## III. SOFT TIMING RECOVERY ALGORITHM

Let us consider the linearly modulated transmitted signal $s(t)$ written as:

$$s(t) = \sum_i a_i h(t - iT), \quad (2)$$

where $a_i$ denotes the zero mean i.i.d. transmitted symbols drawn from a 2 -DBPSK constellation, $h(t)$ is the impulse response of the transmission filter which is according the IEEE 802.15.6 standard a square root raised cosine filter and $T$ is the symbol period.

The received signal is:

$$r(t) = s(t - \tau) + n(t), \quad (3)$$

where $\tau$ is an unknown delay introduced by the channel and $n(t)$ is an additive white Gaussian noise (AWGN) of zero mean and variance $\sigma^2$. In the following paragraph we develop the ML based time delay estimator.

### A. Maximum Likelihood Based Estimator

The time delay is estimated in the ML sense by maximizing the likelihood function according to the following equation [12]:

$$\hat{\tau} = \arg\max_u \Lambda(u, \mathbf{a}), \quad (4)$$

where $\mathbf{a}$ is the vector of the transmitted symbols,

$$\Lambda(u,a) = \exp\left(\frac{1}{2\sigma^2}\int_{T_0} |r(t) - s(t-u)|^2 \, dt\right) \quad (5)$$

is the likelihood function and $T_0$ is the observation interval.

Equivalently, the log-likelihood function $\Lambda_L(u,a)$ can be used instead of $\Lambda(u,a)$. According to [13], when $T_0$ is large enough, the maximum likelihood estimator of the time delay is given by:

$$\hat{\tau} = \arg\max_u \Lambda_L(u, \mathbf{a}), \quad (6)$$

where

$$\Lambda_L(u,\mathbf{a}) = \sum_k \Re\left\{\frac{a_k^* x_k(u)}{\sigma^2}\right\}, \quad (7)$$

where $\Re\{z\}$ is the real part of $z$ and $x_k(u)$ is the matched filter output of the received signal given by:

$$x_k(u) = \sum_i a_i g\big((k-i)T - (\tau - u)\big) + \int_{T_0} h(t - kT - u)n(t)dt \quad (8)$$

$g$ is the convolution of the transmission filter $h$ with its adapted matched filter.

For DPSK modulated signals, one generally uses a differential decoding at the receiver which consists in demodulating $z_k(u) = x_k(u)x_{k-1}^*(u)$ instead of the received signal $x_k(u)$.

Considering $d_k = a_k a_{k-1}^*$, the ML estimation of $\tau$ is equivalent to the maximization with respect to $u$ of the following likelihood function:

$$\Lambda_L^d(u,d) = \sum_k \Re\left\{\frac{d_k^* z_k(u)}{\sigma^2}\right\}. \quad (9)$$

Since it is difficult in practice to solve $\frac{\partial \Lambda_L^d(u,d)}{\partial u} = 0$ with respect to $u$, adaptive algorithms are implemented; their objective is to minimize the derivative of the log-likelihood function $\Lambda_L^d(u,d)$ toward 0 using the following equation:

$$\hat{\tau}_k = \hat{\tau}_{k-1} + \mu e_k(d_k, \hat{\tau}_{k-1}), \quad (10)$$

where $\mu$ is the step size and $e_k(d_k, \hat{\tau}_{k-1})$ is the updating error expressed as:

$$e_k(d_k, \hat{\tau}_{k-1}) = \Re\left\{d_k^* \frac{\partial z_k(u)}{\partial u}\bigg|_{u=\hat{\tau}_{k-1}}\right\}. \quad (11)$$

Such timing error detector is called the Maximum Likelihood Detector (MLD) [14]. The step size can be optimized [15] [16] but this is beyond the scope of our paper. In practice, the term $\frac{\partial z_k(u)}{\partial u}\bigg|_{u=\hat{\tau}_{k-1}}$ can be obtained by interpolating the sampled version of the received signal and evaluating the derivative of the obtained interpolator at $\hat{\tau}_{k-1}$.

The symbols $d_k$ can be known by the receiver only if $a_k$ and $a_{k-1}$ are part of the preamble. The adaptive algorithm is then called a DA or a supervised time recovery detector. However, if $a_k$ and $a_{k-1}$ belong to the PLCP header or the PSDU, they are unknown to the receiver. In this case, we can implement a NDA timing recovery technique which uses a hard estimate $\hat{d}_k$ instead of $d_k$. Nevertheless, this alternative leads to a decrease in the estimation accuracy especially at low SNR where the hard decision is not reliable.

Our contribution proposes a solution in order to deal with this problem. Indeed we can take advantage of the demodulation block by exploiting the soft information coming from of the soft demapper to enhance the performance of the time synchronizer when no reference signals are sent. To summarize, the proposed algorithm consists in a combination of the DA adaptive estimator when the received symbol is a part of the preamble and a soft adaptive estimator when the received signal belongs to the PLCP header or to the PSDU. The soft algorithm for both the $\pi/2-$DBPSK and $\pi/4-$DQPSK mapping techniques is going to be presented in the next two paragraphs.

### B. $\pi/2-$DBPSK modulation

From the classical ML approach, we hereafter derive a new time delay estimation technique which uses the soft information from the demapper.

Let us consider $\lambda_k$ the soft output of the demapper at time index $k$ given by:

$$\lambda_k = \ln\left(\frac{P[b_k=1]}{P[b_k=0]}\right) = \ln\left(\frac{P[d_k=-j]}{P[d_k=j]}\right). \quad (12)$$

As it has been mentioned before, for $\pi/2$-DBPSK mapping $d_k$ can be considered as a binary modulated signal. Based on [17] in which we proposed a soft estimation of BPSK modulated symbols, we propose the following expression of $\tilde{d}_k$, the soft estimation of $d_k$:

$$\tilde{d}_k = j \tanh\left(\frac{\lambda_k}{2} + \frac{\Im\{z_k(\hat{\tau}_{k-1})\}}{\sigma^2}\right), \quad (13)$$

where $\Im\{z\}$ is the imaginary part of $z$.
The updating equation then becomes:

$$\hat{\tau}_k = \hat{\tau}_{k-1} + \mu e_k(\tilde{d}_k, \hat{\tau}_{k-1}), \quad (14)$$

where the soft symbol is given by (13).

### C. $\pi/4-$DQPSK modulation

For a $\pi/4-$DQPSK mapping, $d_k$ takes value from the constellation set $V = \{v_0, v_1, v_2, v_3\}$ where: $v_0 = e^{j\frac{\pi}{4}}$, $v_1 = -v_0$, $v_2 = v_0$ and $v_3 = -v_0$.
Choosing the following mapping:

$v_0 \leftrightarrow 00, v_1 \leftrightarrow 11, v_2 \leftrightarrow 10$ and $v_3 \leftrightarrow 01$

and averaging the likelihood function in (9) on the possible values of $d_k$ leads to:

$$\Lambda(u) = \prod_k \sum_{i=0}^3 P[d_k = v_i] \exp\left(\frac{\Re\{v_i^* z_k(u)\}}{\sigma^2}\right). \quad (15)$$

The expression of $P[d_k = v_i]$ is given by:

$$P[d_k = v_i] = \beta_k \exp\left((2b_{2k}-1)\frac{\lambda_{2k}}{2} + (2b_{2k+1}-1)\frac{\lambda_{2k+1}}{2}\right), (16)$$

where:

$$\lambda_m = \ln\left(\frac{P[b_m=1]}{P[b_m=0]}\right), \quad m \in \{2k, 2k+1\},$$

$$\beta_k = \frac{1}{\cosh\left(\frac{\lambda_{2k}}{2}\right)\cosh\left(\frac{\lambda_{2k+1}}{2}\right)}. \quad (17)$$

By replacing $P[d_k = v_i]$ into (15) with its expression (16) and using (15) we get:

$$\Lambda(u) = \prod_k 4\beta_k \cosh\left(\frac{\lambda_{2k}}{2} + \frac{2\Re\{v_0^*\}\Re\{z_k(u)\}}{\sigma^2}\right) \\ \times \cosh\left(\frac{\lambda_{2k+1}}{2} - \frac{2\Im\{v_0^*\}\Im\{z_k(u)\}}{\sigma^2}\right). \quad (18)$$

Differentiating the log-likelihood function $\Lambda_L(u)$ with respect to $u$ leads to:

$$\frac{\partial \Lambda_L(u)}{\partial u} = \frac{2}{\sigma^2} \sum_K \Re\left\{\tilde{d}_k^* \frac{\partial z_k(u)}{\partial u}\right\}, \quad (19)$$

Where:

$$\Re\{\tilde{d}_k\} = \Re\{v_0^*\}\tanh\left(\frac{\lambda_{2k}}{2} + \frac{2\Re\{v_0^*\}\Re\{z_k(u)\}}{\sigma^2}\right),$$
$$\Im\{\tilde{d}_k\} = \Im\{v_0^*\}\tanh\left(\frac{\lambda_{2k+1}}{2} - \frac{2\Im\{v_0^*\}\Im\{z_k(u)\}}{\sigma^2}\right). \quad (20)$$

In practice, like the $\pi/2-$DBPSK case, we propose to estimate $\tau$, using the updating equation (14) where the real and conjugate parts of $\tilde{d}_k$ are given by (20) for $u = \hat{\tau}_{k-1}$.

As displayed in the next paragraph, the proposed algorithm enhances the performance of the IEEE 802.15.6 NB signal; it is to be highlighted that this improvement is obtained with very low additional complexity. On one side, the soft information is obtained directly from the demapper and not from the channel decoder and thus there is only one direct iteration; on the other side, $\tanh(x)$ is linearized and approximated by $x$ for small values of $x$ and by $+1$ (resp. $-1$) for large positive (resp. large negative) x values.

## IV. SIMULATION RESULTS

The present section displays the performance enhancement brought by the implementation of the proposed soft estimator using the PPDU format given by the IEEE 802.15.6 standard for NB systems.

Let us begin by describing the frame structure utilized in the timing recovery process. The received packet is composed of 201 symbols distributed as follows:
- a 90-bit-preamble-sequence mapped into 90 $\pi/2-$DBPSK symbols. This preamble sequence is known by the receiver.
- a 31-bit-PLCP-header-sequence mapped into 31 $\pi/2-$DBPSK symbols, unknown by the receiver.
- a 160-bit-PSDU-sequence mapped into 80 $\pi/4-$DQPSK symbols also unknown by the receiver.

The 201 symbol block is passed through a square root raised cosine filter with a roll-off factor $\alpha = 0.3$. The filtered signal, in which a time delay $\tau$ is introduced, is transmitted through an AWGN channel and finally matched filtered at the receiver. A soft demapper is placed at the demodulation block which calculates the log-likelihood-ratios (LLR), $\lambda_k$, of the received signal samples. The output, $\lambda_k$, of the soft demapper is injected into the adaptive timing synchronizer to update the current time delay estimate, $\hat{\tau}_k$. For the first 90 received preamble symbols, the time delay estimator operates in the DA mode using equation (10), where $d_k$ is obtained from the transmitted preamble symbols. For the next received symbols, the synchronizer switches to the soft mode using the updating equation (14), where $\tilde{d}_k$ is a soft symbol whose value is computed using the LLR coming from the demapper at time index k. $\tilde{d}_k$ is calculated using (13), if the received sample belongs to the PLCP header, and using (20), if the received sample belongs to the PSDU.

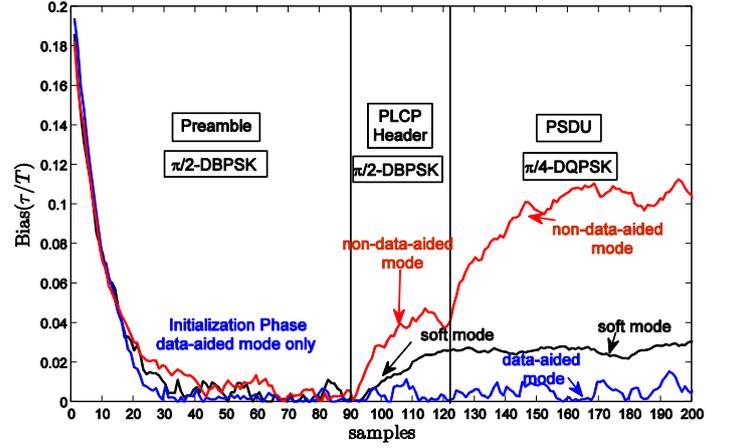

**Fig.3 Bias of the normalized estimated time delay at each sample, SNR=5dB**

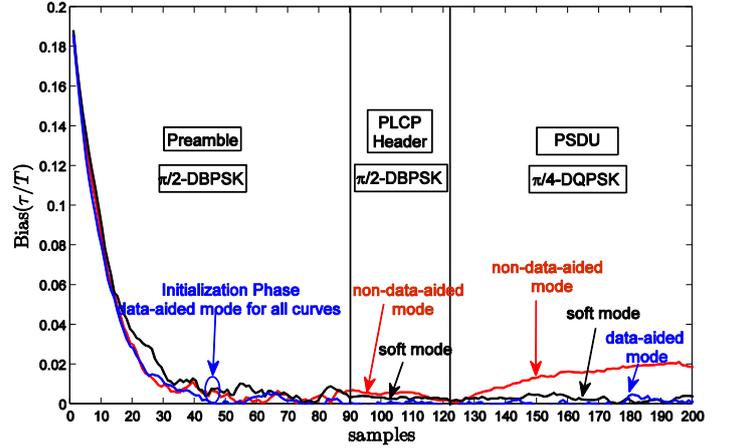

**Fig.4 Bias of the normalized estimated time delay at each sample, SNR=10dB**

In order to evaluate the improvement brought by the soft algorithm, we compare in Fig. 3 and 4 the bias of the normalized time delay estimate, $\tau/T$, at each received sample using the proposed soft technique and the NDA technique. In the NDA approach, $\tau$ is estimated according to (10), where $d_k$ is hardly estimated. These two approaches start after the DA mode, at the beginning of the PLCP header. The SNR is equal to 5 dB in Fig. 3 and to 10 dB in Fig. 4. As it is shown, there is a performance enhancement with the soft technique (black curve) for both $\pi/2-$DBPSK and $\pi/4-$DQPSK mapping in comparison with the NDA technique (red curve) when the SNR is low. For higher SNR values, equivalent performance are observed using the soft and the NDA modes for a $\pi/2-$DBPSK mapping and the improvement appears in the case of a $\pi/4-$DQPSK mapping. The blue curve represents the bias evolution if a DA estimator was implemented and it only serves as a reference. It is obvious that for high SNR values, the soft estimator can achieve equivalent results to the DA one.

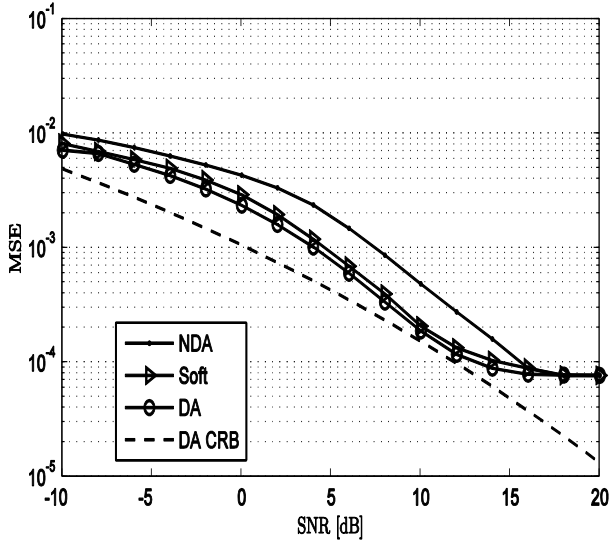

**Fig.5 MSE vs SNR for π/2-DBPSK modulated signal**

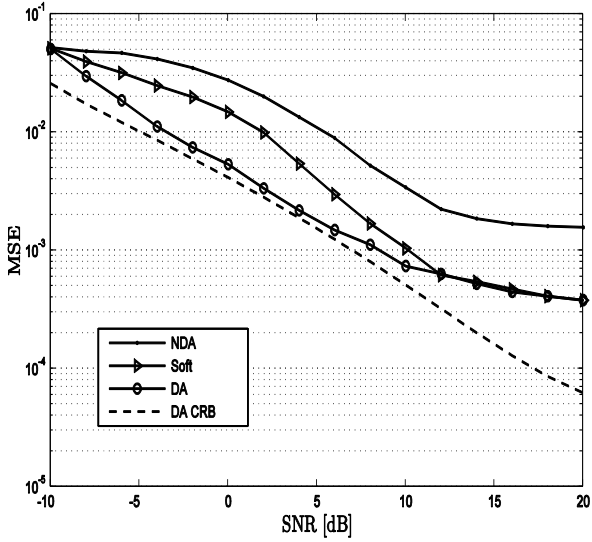

**Fig.6 MSE vs SNR for π/4-DQPSK modulated signal**

In Fig. 5 and 6 we compare the performance of the NDA, soft and DA modes in terms of mean square error (MSE) for a time delay $\tau = 0.1T$ and different SNR values. The MSE of each technique is also compared to the CRB of the DA estimator which is considered as its theoretical lower bound.

Indeed, the CRB verifies $E\left[\left(\frac{\hat{\tau}_k - \tau}{T}\right)^2\right] \geq \mathrm{CRB}\left(\frac{\tau}{T}\right)$ for any unbiased estimator $\hat{\tau}_k$ of $\tau$ [18]. Results are given for 100 symbol blocks averaged over 500 Monte Carlo iterations. $\hat{\tau}_k$ is initialized to 0 and its estimated value is depicted at the end of the block when the steady state is achieved. Fig. 5 (resp. Fig. 6) depicts the performance of the 3 algorithms for a π/2−DBSK (resp. π/4−DQPSK) mapping. In both cases, the soft mode performs better thant the NDA mode and it approaches the DA mode performance over a large interval of SNR values. The MSE of all the algorithms approaches the CRB and this confirms the consistence of the proposed estimator. Numerical
results are presented in TABLE IV and V.

**Table IV. MSE VALUES FOR A π/2-DBPSK MAPPING**

| SNR [dB] | DA | Soft | NDA |
|---|---|---|---|
| 0 | $2.5 \times 10^{-3}$ | $3 \times 10^{-3}$ | $4 \times 10^{-3}$ |
| 10 | $1.9 \times 10^{-4}$ | $2.1 \times 10^{-4}$ | $5 \times 10^{-4}$ |

**Table V. MSE VALUES FOR A π/4-DQPSK MAPPING**

| SNR [dB] | DA | Soft | NDA |
|---|---|---|---|
| 0 | $5 \times 10^{-3}$ | $1.5 \times 10^{-2}$ | $2 \times 10^{-2}$ |
| 10 | $7 \times 10^{-4}$ | $10^{-3}$ | $4 \times 10^{-3}$ |

The saturation of the MSE at the right side of Fig. 5 and 6 is due to the self-noise of the updating error (11). Similar results are obtained for other roll-off factors.

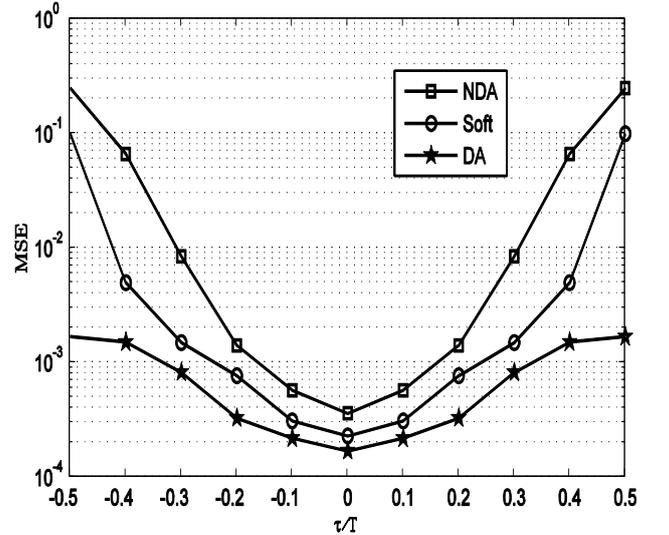

**Fig.7 MSE vs $\tau/T$ for π/2-DBPSK modulated signal, SNR=10dB**

A similar result is shown in Fig. 7 where the MSE of the 3 techniques is plotted as a function of the normalized time delay, $\tau/T$, for a π/2−DBSK modulated signals and a SNR which is equal to 10 dB. Amelioration is observed with the soft technique with respect to the NDA mode for any time delay value. Some numerical results are given in TABLE VI.

**Table VI. MSE VALUES FOR A π/2-DBPSK MAPPING**

| τ / T | DA | Soft | NDA |
|---|---|---|---|
| 0.1 | $2 \times 10^{-4}$ | $3 \times 10^{-4}$ | $6 \times 10^{-4}$ |
| 0.3 | $8.5 \times 10^{-4}$ | $1.5 \times 10^{-3}$ | $9 \times 10^{-3}$ |

## V. CONCLUSION

In this paper, we presented a time delay recovery algorithm based on the ML estimator for the NB IEEE 802.15.6 WBAN. The proposed technique combines a DA with a soft synchronizer using soft information from the demapper block to provide accurate estimates of the transmitted symbols. The proposed soft estimator has shown better synchronization performance in comparison with the NDA estimator when no pilot symbols are sent and can reach the DA mode performance over a large interval of SNR values. The relevance of the so proposed time synchronization processing offers the possibility to reduce power consumption and increase spectrum efficiency while enhancing the performance compared to the NDA scheme through offering the possibility of pilot signals shortening. To measure the performance of our estimator, we must compare them to Cramer-Rao bounds [19] [20] [8] of relevance [21]. Some future works aim at transposing our approach to the Bayesian canvas [22] so as to reach higher data rate [23].

**Acknowledgement**

This work has been supported by the franco-chinese NSFC-ANR program under the Greencocom Project.

## Biographies

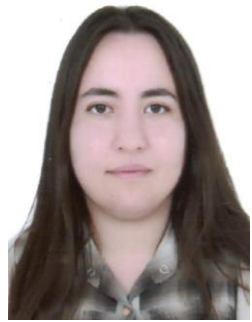

*Imen Nasr,* is a PhD student in Information and Communications Technologies in both the Ecole Nationale Supérieure de Techniques Avancées (ENSTA) ParisTech, France and the Higher School of Communications of Tunis (Sup'Com), Tunisia. She received an Engineering and a Master Degree in Telecommunication from Sup'Com, University of Carthage, Tunisia, in 2011. Her research activities include synchronization and localization for Wireless Body Area Networks (WBAN).

Email: imen.nasr@ensta-paristech.fr

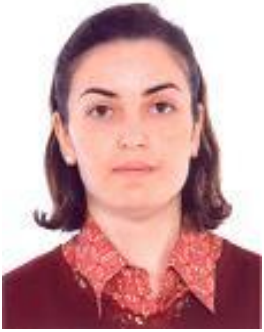

***Leïla Najjar Atallah***, is an associate Professor in the Higher School of Communications of Tunis (Sup'Com), with research activities in the research laboratory Communications, Signal and Image (COSIM) in Sup'Com. She received an engineering degree from Polytechnic School of Tunisia, a Master Degree in Automatic and Signal Processing from Ecole Supérieure d'Electricité Supélec, and a PhD in Sciences from University Paris XI. Her current research is in signal processing for wireless communications, including channel estimation, synchronization and localization. She is also interested in sparse regularization problems and statistical signal processing.
Email: leila.najjar@supcom.rnu.tn

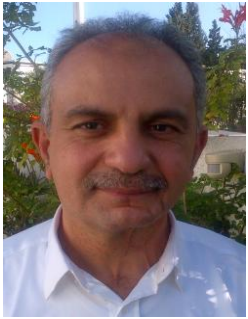

***Sofiane Cherif,*** is professor in telecommunication engineering at the Higher School of Communications of Tunis (SUP'COM), University of Carthage. He received engineering, MS degrees, and Ph.D. in Electrical Engineering from the "National Engineering School of Tunis (ENIT) ", University of Tunis-El Manar, in 1990 and 1998, respectively and the "Habilitation Universitaire" in Telecommunication from SUP'COM, in 2007. From 2011 to 2014, he was the head of the doctoral school in ICT, and presently, he is the head of COSIM research Lab at SUP'COM. His current research interests are signal processing for communications, resource allocation, interference mitigation in wireless networks, wireless sensor networks and cognitive radio.
Email: sofiane.cherif@supcom.rnu.tn

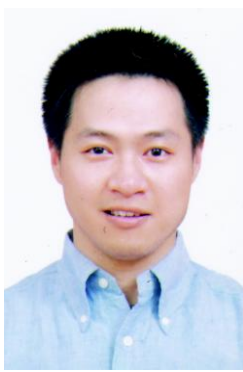

***Jianxiao Yang***, is a research engineer at Lab UIIS of ENSTA PARIS-TECH (2013.10-Now). He received a PhD degree from Zhejiang University, China in 2007 on Information and Communication Engineering. He joined the research Lab SATIE of the Ecole Normale Superieure de Cachan (2007.11-2009.05), then the Lab UEI of ENSTA PARIS-TECH (2009.06-2009.09) and the Lab-STICC of Telecom Bretagne (2009.10-2013.9).. His research interests include Error Control Coding, Synchronization, Channel Estimation, MIMO Detection and interference cancellation. He has a strong experience in Wireless Digital Communication and Broadcasting systems.
Email: jianxio.yang@ensta-paristech.fr

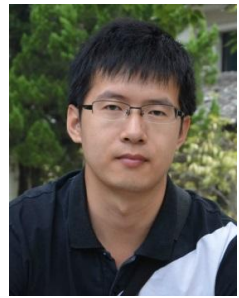

***Kunlun Wang,*** is a PhD student at Network Coding and Transmission Laboratory, Shanghai Jiao Tong University, Shanghai, China. He received the B.S. degree from Hangzhou Dianzi University, Hangzhou, China, in 2009, and the M.S. degree from South China University of Technology, Guangzhou, China, in 2012. His research interests include energy efficient communications, cross-layer design and MIMO systems.
Email: kunlun1228@sjtu.edu.cn